\documentclass[final,hidelinks,onefignum,onetabnum]{siamart250211}

\usepackage{amsfonts}
\usepackage{subfig}
\usepackage{amsmath}
\usepackage{amssymb}
\usepackage{booktabs}
\usepackage{bm}
\usepackage{hyperref}
\usepackage{tcolorbox}
\usepackage{natbib}

\hypersetup{colorlinks,citecolor=blue}

\newcommand*\diff{\mathop{}\!\kern0pt\mathrm{d}}
\newcommand{\iu}{\mathrm{i}\mkern1mu}

\ifpdf
\hypersetup{
	pdftitle={NUFFT for the Fast COS Method},
	pdfauthor={Fabien Le Floc'h}
}
\fi
\title{NUFFT for the Fast COS Method\thanks{Date: July 14, 2025.}}
\author{Fabien Le Floc'h\\ \email{fabien@2ipi.com}}
\date{July 14, 2025}

\begin{document}
	\maketitle
	\begin{abstract}
The COS method is a very efficient way to compute European option prices under L\'evy models or affine stochastic volatility models, based on a Fourier Cosine expansion of the density, involving the characteristic function. This note shows how to compute the COS method formula with a non-uniform fast Fourier transform, thus allowing to price many options of the same maturity but different strikes at an unprecedented speed.
	\end{abstract}
	\section{Introduction}
For a finite number of given  coefficients ${f}_k \in \mathbb{C}$, $k \in I_N$, the \emph{direct} one dimensional non-uniform discrete Fourier transform (NUDFT)
 reads \citep{keiner2007nfft,knopp2023nfft}
\begin{equation}
\hat{f}_j = \sum_{k \in I_N} f_k e^{-2\pi\iu  k x_j}	\,,\label{eqn:nufft_type1}
\end{equation}
where $(x_j)_{j=1,...,J} \in [-1/2, 1/2)$ are sampling points and $I_N = \left\{ -\frac{N}{2}, -\frac{N}{2}+1, ..., \frac{N}{2}-1\right\}$ for even $N$, $I_N = \left\{ -\frac{N-1}{2}, -\frac{N-1}{2}+1, ..., \frac{N-1}{2}\right\}$ for odd $N$. It maps an equidistant sampling frequency $k$ to a non-equidistant space $x_j$.
This transformation is also known as \emph{type-2} NUDFT.

The \emph{adjoint} or \emph{type-1} NUDFT maps a non-equidistant domain to an equidistant one and reads
\begin{equation}
y_k = \sum_{j=1}^J \hat{f}_j e^{2\pi\iu  k x_j}	\,,
\end{equation}
for $k \in I_N$. It is in general not the inverse of the forward NUDFT and $y_k$ is not the same as $f_k$.

The NNDFT or \emph{type-3} NUDFT involves non-equidistant sampling point in both frequency and spatial domains:
\begin{equation}
\hat{f}_j = \sum_{k \in 0}^{N-1} {f}_k e^{-2\pi\iu  v_k x_j}	\,,
\end{equation}
for $j=1,...,J$.

NUDFTs can be computed very efficiently by one or multiple applications of the fast Fourier transform algorithm, leading to so-called NUFFT methods.
For example, the type-1 and type-2 are typically computed via a resampling correction in equidistant domain, a fast Fourier transform and a resampling to map from equidistant to non-equidistant domain.

\citet{andersen2022high} introduced the use of NUFFT to compute European option prices under stochastic volatility models through their characteristic function. They propose three different methods: \begin{itemize}
\item The so-called Parseval method, which consists in a single integration of the characteristic function multiplied by a payoff dependent function and a damping parameter $\alpha$. The technique is well known (see \citet{carr1999option,lord2006optimal,schmelzle2010option,floc2014fourier}), although in the previous literature, caching of the characteristic function was used to speed up the computation.
\item The density method, where distinct integrations are performed: one to compute the density from the characteristic function at a set of relevant nodes, and one over the density and the payoff function to compute the option price. NUFFT allows this method to perform adequatly, and makes this method practical.
\item The CDF method, which is very similar to the density method, except that two integrals are computed over the cumulative distribution function calculated from the characteristic function.
\end{itemize}
All three involve the use of a type-3 NUFFT.
\citet{andersen2022high} present timing results for the methods, showing that the Parseval method significantly outperforms fast techniques such as the COS method of \citet{fang2009novel} on the problem of pricing many vanilla European options of different strikes, for a given maturity.

In this note, we show that the COS method can be adapted to make use of the simpler \emph{type-2} NUFFT, thus unleashing unprecedented speed for pricing options.

\section{The Classic COS Formula}
We consider an asset $F$ with a known (normalized) characteristic function
\begin{equation}
\phi(x) = \mathbb{E}\left[e^{i x \ln \frac{F(T,T)}{F(0,T)}}\right]\,.
\end{equation}
$F(0,T)$ is typically the forward price to maturity $T$ of an underlying asset $S$. For example, for an equity with spot price $S$, dividend rate $q$ and interest rate $r$, we have $F(0,T)=S(0)e^{(r-q)T}$. The price of a Put option with the COS method reads

\begin{align}
P(K,T) &= B(T) K\left[\frac{1}{2}\Re\left(\phi(0)\right)U_0^{\textmd{Put}}+ \sum_{k=1}^{M-1} \Re\left(\phi\left(\frac{k\pi}{b-a}\right)U_k^{\textmd{Put}} e^{ik\pi\frac{-x-a}{b-a}}\right) \right]\label{eqn:cos_method}
\end{align}
with $U_0^{\textmd{Put}} =\frac{2}{b-a}\left(e^a-1-a\right)$ and for $k \geq 1$
\begin{align}
U_k^{\textmd{Put}} &=\frac{2}{b-a} \left[ \frac{1}{1+\eta_k^2}\left( e^a + \eta_k \sin\left(\eta_k a\right) - \cos\left(\eta_k a\right)  \right) - \frac{1}{\eta_k} \sin\left(\eta_k a\right) \right] 
\end{align}
where $B(T)$ is the discount factor to maturity, $x = \ln \frac{K}{F(0,T)}$ and $\eta_k=\frac{k\pi}{b-a}$.

The truncation range $[a, b]$ is commonly chosen according to the first two or four cumulants $c_1$ and $c_2$ of the model considered  using the rule \begin{align*}a = c_1 - L \sqrt{|c_2|+\sqrt{|c_4|}} &\,,\quad b = c_1 + L \sqrt{|c_2|+\sqrt{|c_4|}}\,, \end{align*} with $L$ is a truncation level. This rule is relatively robust for different models, parameters and option maturities.
Recently, \citet{junike2024number} proposed another way\footnote{With several caveats: this does not work for all models, for example it is not applicable to the variance gamma model, and the estimate of $M$ requires a different non-trivial numerical integration.} to find the truncation range and the associated number of terms $M$ to guarantee a maximum error tolerance.

The Call option price is obtained through the Put-Call parity relationship. Digital options or the probability density may be computed using different payoff coefficients $U_k$.

\section{Rewriting the COS Method for NUFFT}
We consider the evaluation of the COS method formula for a set of $J$ strikes $K_j$, $j=1,...,J$ at a given maturity $T$. 

We want to rewrite Equation \ref{eqn:cos_method} in the form of Equation \ref{eqn:nufft_type1}. We have
\begin{align*}
	&\sum_{k=1}^{M-1} \Re\left(\phi\left(\frac{k\pi}{b-a}\right)U_k^{\textmd{Put}} e^{ik\pi\frac{-x-a}{b-a}}\right)
	= \Re\left(\sum_{k=1}^{M-1} \phi\left(\frac{k\pi}{b-a}\right)e^{-ik\pi\frac{a}{b-a}} U_k^{\textmd{Put}} e^{-2ik\pi\frac{x}{2(b-a)}}\right)\,.
\end{align*}
We know that the log-moneynesses we will evaluate obey $x \in (a,b)$ and in fact they must be sufficiently far away from the boundaries in order to preserve the accuracy of the COS method. We thus have $x_j \in (\frac{a}{2(b-a)}, \frac{b}{2 (b-a)}) \in \left[-\frac{1}{2},\frac{1}{2}\right)$. Let $N=2M$, we may thus define 
\begin{subequations}
\begin{align}
	f_k &= \phi\left(\frac{k\pi}{b-a}\right)e^{-ik\pi\frac{a}{b-a}} U_k^{\textmd{Put}}\,,\quad\textmd{ for } k=1,...,N/2-1\\
	f_k &= 0\,,\quad\textmd{ for } k=-N/2,...,-1\,.\\
	f_0 &= \frac{1}{2}\phi(0)U_0^{\textmd{Put}}\,.\\
	x_j &= \frac{1}{2(b-a)}\ln \frac{K_j}{F(0,T)}\,,\quad\textmd{ for } j=1,...,J\,.
\end{align}
\end{subequations}
and compute the type-2 NUFFT $\hat{f}_j$ at the points $x_j$. Then the Put option prices read
\begin{align}
P(K_j,T) &= B(T) K_j \Re\hat{f_j}
\end{align}

\section{Rewriting the Alternative COS Method for NUFFT}
Although slightly more involved, it is also possible to use the type-2 NUFFT for the alternative COS formula of \citet{floc2020more}:
\begin{equation}
	P(K,T) = B(T) \left[\frac{1}{2}\Re\left(\phi(0)\right)V_0^{\textsf{Put}}(x)+ \sum_{k=1}^{M-1} \Re\left(\phi\left(\frac{k\pi}{b-a}\right) e^{-\iu k \pi\frac{a}{b-a}}\right)V_k^{\textsf{Put}}(x) \right]\,.\label{eqn:cos_f_price}
\end{equation}
with
\begin{align}
	V_0^{\textsf{Put}}(x)&=2F\frac{e^a-e^x+e^x(x-a)}{b-a}\,,\\
	V_k^{\textsf{Put}}(x)&=  \frac{2F}{(b-a)\left(1+\eta_k^2\right)}\left[e^a - \cos\left(\eta_k(x-a)\right)e^x - \eta_k\sin\left(\eta_k(z-a)\right)e^x  \right]\nonumber\\
	&\quad+ \frac{
	2F}{(b-a)\eta_k} \sin\left(\eta_k(x-a)\right)e^x \quad \textmd{ for } k=1,...,M-1\,.\label{eqn:cos_f_vk}
	\end{align}
We need to split the dependency on $x$ in $V_k$ and rewrite the $\cos$ and $\sin$ to use the exponential form. This leads to 
\begin{subequations}
	\begin{align*}
		f_k &= \phi\left(\frac{k\pi}{b-a}\right)e^{-\frac{ik\pi a}{b-a}} \left[ \frac{-1 - \iu \eta_k}{(b-a)(1+\eta_k^2)} + \frac{\iu}{(b-a)\eta_k}\right]\textmd{ for } k=1,...,\frac{N}{2}-1\,,\\
		f_k &= \phi\left(\frac{-k\pi}{b-a}\right)e^{\frac{ik\pi a}{b-a}} \left[ \frac{-1 + \iu \eta_{-k}}{(b-a)(1+\eta_{-k}^2)} - \frac{\iu}{(b-a)\eta_{-k}}\right]\textmd{ for } k=-\frac{N}{2}-1,...,-1\,,\\		
		f_{-N/2} &= 0\,, \quad f_0 = 0\,,\\
		x_j &= \frac{1}{2(b-a)}\left(\ln \frac{K_j}{F(0,T)} - a \right)\textmd{ for } j=1,...,J\,.
	\end{align*}
	\end{subequations}
	Then the Put option prices read
\begin{align}
P(K_j,T) &= B(T) \Re\left[ K_j\phi(0)\left(2x_j - \frac{1}{b-a}\right) + K_j\hat{f_j} \right. \nonumber\\ 
&\left.+F(0,T)\frac{e^a}{b-a}\phi(0)+F(0,T)\sum_{k=1}^{M-1} \phi\left(\frac{k\pi}{b-a}\right) e^{-\frac{ik\pi a}{b-a}} \frac{2e^a}{(b-a)(1+\eta_k^2)} \right]\,.
\end{align}

We have reduced the range of $x_j$ to $[0,1/2)$, another possibility is to factor out $-a/(b-a)$ to use the full $[-1/2,1/2)$ range.

The interest of this alternative formula is to be more accurate when the strikes are close to the boundaries of the truncation range. 

\section{Numerical Examples}
In our numerical tests, we will make use of the package NFFT for the \emph{Julia} programming language \citep{knopp2023nfft}.

\subsection{Variance Gamma Model}
In order to provide a comparison with the results of \citet{andersen2022high}, we first consider the variance gamma model with parameters defined in Table \ref{tbl:vg_cases}. A minor difference lies in the interest rate $r$: we use the one as defined in the original data from \citet{crisostomo2018speed}, while \citet{andersen2022high} use zero interest rates.

 The characteristic function for the normalized log asset price is given by
\begin{equation}
	\phi(z) = e^{ -\frac{T}{\nu}\ln\left(1 - \iu z \nu \left( \theta + \frac{\sigma^2 \iu z}{2}\right)\right) } e^{\iu z \frac{T}{\nu} \ln\left(1 - \theta \nu - \frac{\sigma^2 \nu}{2}\right)}\,,\label{eqn:vg_char}
\end{equation}
The second exponential term in Equation \ref{eqn:vg_char} is there to ensure martingality of the forward price.

\begin{table}[H]
\centering{
\caption{Variance Gamma: Test Cases from \cite[Table 6]{andersen2022high}. The spot price $S(0)=100$.\label{tbl:vg_cases}}
\begin{tabular}{lccccccc}
\toprule
Case & $T$ & $\theta$ & $\nu$ & $\sigma$ & $r$ & $\frac{2T}{\nu}$ & PDF at origin \\
\midrule
1 & 1.0 & -0.1436 & 0.3 & 0.12136 & 0.1 &6.66667 & Smooth \\
2 & 0.1 & -0.1436 & 0.3 & 0.12136 & 0.1 & 0.66667 & Algebraic blow-up \\
4 & 1.0 & 1.5 & 0.2 & 1.0 & 0.02 & 10.0 &  Smooth \\
5 & 0.1 & 1.5 & 0.2 & 1.0 & 0.02 &1.0 & Logarithmic blow-up \\
\bottomrule
\end{tabular}	}
\end{table}

Contrary to what is suggested in \citep{andersen2022high}, Case 4 and 5 were not found to be intractable or inaccurate with the COS method. The divergence they observed may be due to the choice of a too narrow truncation range. We do not reproduce the issue highlighted in \citep[Figure 4]{crisostomo2018speed} and reach\footnote{The reference values in \citep{crisostomo2018speed} are given with 12 digits.} accuracies below $10^{-12}$  with $M=2^{20}$ and $L=20$ for both $T=0.1$ and $T=1.0$.  In fact, with $M=2^{10}$ points and $L=10$ we already obtain a maximum absolute error below $10^{-12}$ for $T=1$ and below $3 \cdot 10^{-5}$ for $T=0.1$.

Case 2 requires a very large number of terms $M$ to achieve high accuracy.

We report in Table \ref{tbl:vg_nufft} the number of options priced per second for the classic COS method and the NUFFT COS method for batches of equidistant\footnote{\citet{andersen2022high} measures the throughput on batches of uniform log-strikes, which may help make the NUFFT more efficient, since the space coordinate is proportional to log-strikes. In our test it did not appear to make a difference for the COS method.} strikes  $K \in [60, 140]$.
\begin{table}[H]
	\caption{Number of options priced per second. Case 1 uses $M=128$ and $L=10$ and Cases 2 and 5 use $M=1024$ and $L=10$ to reach a maximum absolute error below $10^{-4}$.\label{tbl:vg_nufft}}
	\centering{
	\begin{tabular}{llrrrrr}
	\toprule
Case & Method & \multicolumn{5}{c}{Number of strikes}\\\cmidrule(lr){3-7}
& & 10 & 25 & 100 & 500 & 2500 \\ \midrule
1 & Classic & 412k & 583k & 728k & 776k & 797k  \\ 
& NUFFT &  130k & 324 & 1245k & 5254k &14874k  \\
2 and 5 & Classic &  48k & 72k & 96k & 105k & 107k \\
				& NUFFT & 43k & 107k & 422k & 1988k & 7795k \\  	  
	\bottomrule
	\end{tabular}}
	\end{table}
As expected, the NUFFT implementation is much faster for large number of strikes, while the classic COS method reaches a threshold above 100 strikes. The classic COS method is however faster for small number of strikes, as the overhead of the NUFFT implementation is not negligible then. The NUFFT also performs better as the number of terms $M$ used in the sum increases: Cases 2 and 5 use $M=1024$ while Case 1 uses $M=128$ in order to reach a maximum absolute error below $10^{-4}$.

\subsection{Heston Model}
In the Variance Gamma model, the characteristic function is relatively fast to evaluate, but this model is not so popular in practice. We thus consider the Heston model, which is widely used in practice. The normalized characteristic function is given by
\begin{equation}
	\phi(z) = e^{\frac{v_0}{\sigma^2}\frac{1-e^{-DT}}{1-Ge^{-DT}}(\kappa-\iu\rho\sigma z -D) + \frac{\kappa\theta}{\sigma^2}\left( (\kappa-\iu\rho\sigma z -D)T - 2 \ln\left(\frac{1-Ge^{-DT}}{1-G}\right)\right)}
\end{equation}
with 
\begin{align}\label{eqn:DandG}
	\beta = \kappa- \iu\rho\sigma z,\,,\quad D = \sqrt{(\beta^2+(z^2+\iu z)\sigma^2}\,,\quad	G =\frac{\beta-D}{\beta +D}\,.
\end{align}

We use the parameters from \citet{floc2014fourier}, that is $\kappa=1$, $\theta=0.1$, $\sigma= 1$, $v_0=0.1$, $\rho=-0.9$ and $T=2$ years with $r=0\%$. 
\begin{table}[H]
	\caption{Number of options priced per second for a batch of uniform strikes under the Heston model, with $L=8$ and various number of points $M$.\label{tbl:heston_nufft}}
	\centering{
	\begin{tabular}{llrrrrrrr}
	\toprule
	 $M$ & Method  & \multicolumn{5}{c}{Number of strikes} & RMSE & MAE\\\cmidrule(lr){3-7}
& & 10 & 25 & 100 & 500 & 2500 & \\ \midrule
256 & Classic & 147k & 248k & 371k & 430k & 444k  & 5.62e-06 & 1.31e-05 \\ 
& NUFFT &  84k & 208k & 812k & 3625k & 11839k  & 5.62e-06 & 1.31e-05\\
1024 & Classic &  36k & 59k & 87k & 99k & 102k & 3.07e-10 & 6.06e-10\\
				& NUFFT & 31k & 78k & 309k & 1475k & 6071k  & 3.16e-10 & 1.15e-09\\  	  
	\bottomrule
	\end{tabular}}
	\end{table}
The characteristic function is slower to evaluate, and as a consequence, the threshold where the NUFFT implementation becomes faster than the classic implementation is lower (Table \ref{tbl:heston_nufft}). There is a small difference in the root mean square error (RMSE) and the mean absolute error (MAE) between the two methods when $M=1024$: we used the default relative tolerance ($10^{-9}$) for the NFFT package. If we use a lower tolerance of $10^{-16}$, the results become identical. The throughput is then reduced by 20\%.

\begin{figure}
\centering
\includegraphics[width=0.8\textwidth]{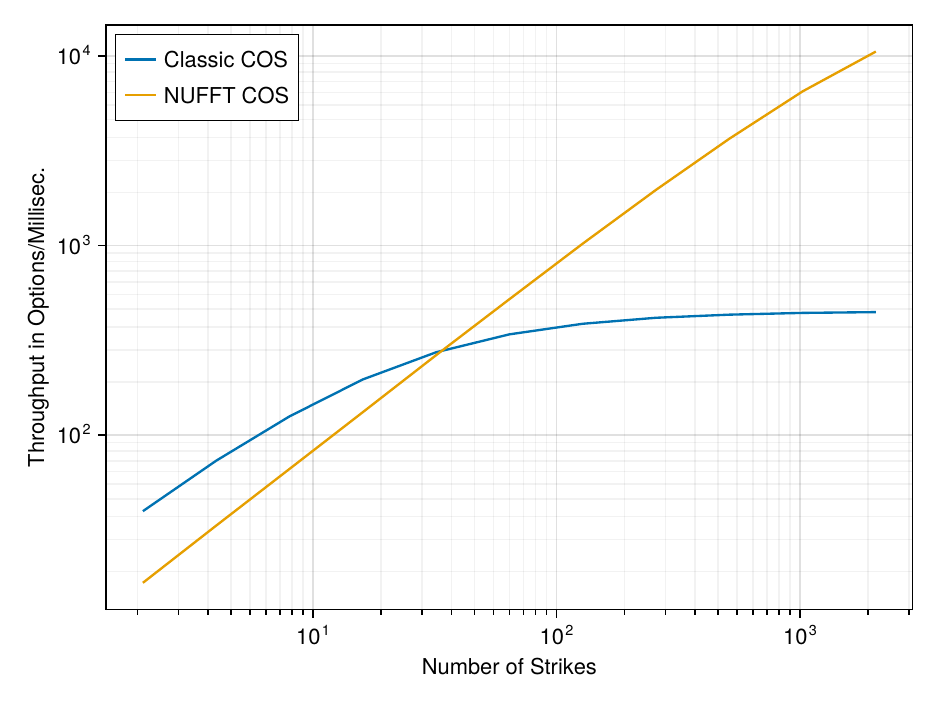}
\caption{Number of options priced per millisecond for the Heston model with $M=256$ and $L=8$ as a function of the number of strikes per batch.\label{fig:heston_nufft}}
\end{figure}

	\section{Conclusion}
	The COS method can make use of the type-2 NUFFT. This leads to a very significant speedup when many options of the same maturity but different strikes are priced. The threshold is around 100 strikes, and depends on the cost of the characteristic function evaluation. When the characteristic function is slow to evaluate, the overhead of the NUFFT implementation is negligible and the method is competitive also for a smaller number of strikes. It also depends on the number of terms used in the COS method: the larger $M$, the more efficient the NUFFT implementation becomes.

	Model calibration rarely makes use of a large number of strikes per maturity. The NUFFT technique however opens up the possibility to compute the probability density at many points very quickly.
\bibliographystyle{plainnat}
\bibliography{cos_nufft}

\end{document}